# An Efficient Recommendation System in E-commerce using Passer learning optimization based on Bi-LSTM

**Hemn Barzan Abdalla [1*], Mehdi Gheisari [2], Awder Ahmed [3], Bahtiyar Mehmed [4], Maryam Cheraghy [5], Yang Liu [6]**

[1,5] School of Computer Science, Wenzhou-Kean University, Wenzhou, China
[1,5] School of Computer Science and Technology, Kean University, Union, NJ, USA
[2] Institute of Artificial Intelligence, Shaoxing University, Shaoxing, Zhejiang, China
[3] Technical College of Engineering, Sulaimani Polytechnic University, Sulaymaniyah, Iraq
[4] School of Economics, Neusoft Institute Guangdong, Foshan, China
[6] Department of Computer Science, Swansea University, Swansea, UK;
[*] Corresponding author: Hemn Barzan Abdalla; habdalla@kean.edu

## Abstract

Online reviews play a crucial role in shaping consumer decisions, especially in the context of e-commerce. However, the quality and reliability of these reviews can vary significantly. Some reviews contain misleading or unhelpful information, such as advertisements, fake content, or irrelevant details. These issues pose significant challenges for recommendation systems, which rely on user-generated reviews to provide personalized suggestions. This article introduces a recommendation system based on Passer Learning Optimization-enhanced Bi-LSTM classifier applicable to e-commerce recommendation systems with improved accuracy and efficiency compared to state-of-the-art models. It achieves as low as 1.24% MSE on the baby dataset. This lifts it as high as 88.58%. Besides, there is also robust performance of the system on digital music and patio lawn garden datasets at F1 of 88.46% and 92.51%, correspondingly. These results, made possible by advanced graph embedding for effective knowledge extraction and fine-tuning of classifier parameters, establish the suitability of the proposed model in various e-commerce environments.

**Keywords:** Recommendation system, E-commerce, Passer Learning optimization, Bi-LSTM classifier, TF-IDF.

## 1. Introduction

The online shopping has become very common because of the increased use of the Internet and mobile phones in the recent years [1][34][2][5]. This growth has nurtured the emergence of electronic commerce, which has granted people the ability to buy, sell, and barter goods and services based on Internet-enabled platforms [3][4][6]. In e-commerce, understanding consumer behavior is considered the cornerstone of preference, decision, and interaction on the platforms. Recommender systems help drive this ecosystem by leveraging user data on browsing history, search queries, and purchase patterns to offer suggestions for personalization that increase customer satisfaction and drive conversions.

E-commerce encompasses a wide variety of business models, including B2C sites (like Amazon) and C2C marketplaces (such as eBay), and monetization approaches that leverage advertising and subscriptions in addition to freemium models. These areas of strategy — from pricing to supply chain optimization to data security — are keys to thriving and growing in a global economy. Recommendation systems form the core of solving key challenges around customer engagement, conversion optimization, and product relevance. These systems study user behavior and make the shopping experience better, optimize conversions, and personalize products. While impactful, several recommendation approaches exist that involve a number of trade-offs: for example, fuzzy logic-based systems improve diversity at the cost of confidence [8]; hybrid models that combine collaborative and content-based filtering address the issue of accuracy but not cold starts [11]. More advanced scalable implementations include deep learning based [9], cross-domain matrix factorization [12], and self-complementary collaborative filtering [14]. Nevertheless, all of these have challenges, including high computational complexity and limited data generalization [29].

This study aims to address cold-start and scalability issues in e-commerce recommendation systems by integrating Passer Learning Optimization with Bi-LSTM architecture. Moreover, combining with Passer learning optimizers a fusion of TLBO + SSA can help in ensuring enhanced accuracy of recommendations, scalability and convergence with tracking data nature and user preference compliance. This Collaboration-based Bi-LSTM classifier can present multiple content- and collaboration-based feature extraction strategies to limit cold starts and ensure accurate recommendation in low user history states. Developed innovations target towards natural, scalable, and implementable method of recommending products in an e-commerce platform [1].

The following are the main contributions of this research:

- Passer learning optimization (PLO): The research introduces the Passer learning optimization algorithm, a hybrid approach that merges the automatic learning capacity of a teacher with sparrow-inspired principles. The foraging behavior, the producers' guiding behavior, and the sparrow's alarm-signaling behavior are merged to find the optimal solution. This distinctive blend improves convergence rates and performance, differentiating it from conventional optimization techniques.

➢PL-optimized Bi-LSTM is a refined version of the Bi-LSTM architecture. It offers enhanced sequence modeling capabilities, making it particularly valuable for tasks involving sequential data, such as natural language processing and time series analysis. The optimizations introduced contribute to improved performance, including accuracy and convergence speed, improving e-commerce through better product review analysis and personalized recommendations.

This paper is organized as follows: The proposed methodology for the e-commerce recommendation system is described in the second section and the Passer learning optimization technique and related mathematical models are discussed in the third section. The experimental findings of the proposed method are presented in the fourth section, and the conclusion and suggestions for future work comprise the fifth section.

## 2. Literature Review

The previous model's recommendation system in e-commerce is discussed in the below section, and A new fuzzy logic-based product suggestion system was developed by [33] that utilizes users' present interests to anticipate which products are the most pertinent to them when they purchase online. This approach raised the diversity and product rating score but decreased the recommendation system's level of confidence. A deep learning and distributed expression-based application solution for e-commerce product advertising recommendations were created by [9]. This approach had great flexibility and minimized calculating complexity, but it was exceedingly expensive and time-consuming. [10] proposed a heterogeneous product system by combining user, item, and interaction information. This strategy increased quality and decreased rating prediction error but converged too slowly. Sunny Sharma et al. introduced a hybrid recommendation system that anticipates recommendations [11]. The suggested method combined collaborative filtering with content-based filtering. Although this method's recommendations were more accurate, it had a cold start issue. In the cross-domain situation of user overlap in an e-commerce system [12] created a domain matrix factorization that predicted the rating of a product for a frequent user with a fixed cold start issue. Although this method's convergence is excessively sluggish and does not predict all testing data, it does minimize the time complexity. [13] presented the session similarity-based strategy to address the issue of cold-start sessions in the e-commerce area, where no interactive objects in the sessions can aid in discovering users' preferences. This approach lessened the effects of the cold-start movie issue. Self-complementary collaborative filtering is a framework proposed by [14] and can make recommendations in real time using both global and local information. This technique lowered the size of the model and helped with overfitting, but it is time-consuming and cannot be used in large-scale applications. [15] presented a novel strategy that is effectively a hybrid probabilistic matrix model. Although this method had a high computational complexity and a high computing cost, it minimized errors.

## 3. Materials and Methods

The following is a list of the challenges encountered during the current research:

- Improved Accuracy and Efficiency: Our PL-optimized Bi-LSTM model will overcome the shortcomings of the available techniques in reducing computational complexity and enhancing the convergence speed, hence finding applications in real-time dynamic e-commerce environments.
- Performance Improvement: The performance of the integration of Passer Learning Optimization with Bi-LSTM outperforms others, reducing MSE to 1.24%, and increasing F1 scores to 88.58% on key datasets, outperforming traditional models.
- Real-Time Capabilities: Our approach provides the rapidity of decision-making demanded by real-time applications by optimizing computational efficiency-a common weakness with systems involving heterogeneous products.
- Cost-Effectiveness: The PL-optimized Bi-LSTM slashes deep learning systems' time and money, hence very well viable on a large-scale e-commerce platform without hurting recommendation quality.
- Algorithm Efficiency: The proposed model will maintain high user satisfaction while enhancing algorithm efficiency, balancing context-aware personalization with demands for scalability and performance.

The materials and methods used for the recommendation system in e-commerce are described next.

### 3.1 Proposed Methodology for Recommendation System in E-commerce

This research primarily aims to create an e-commerce recommendation system employing a Passer learning optimization-based Bi-LSTM classifier. Initially, the recommendation dataset [31] is input, and the data are pre-processed to enhance quality. The pre-processing phase includes stemming, lemmatization, and tokenization processes, improving data quality. The pre-processed data is forwarded to content- and collaboration-based feature extraction, where the TF-IDF and the graph embedding-enabled features are extracted. The graph embedding enables extraction of the data features in the form of a discrete graph which is represented in vector form, and the TF-IDF features measure the relationship between Word and the document. Extracting these features reduces the data's dimensionality by eliminating irrelevant features and it reduces data dimensionality by eliminating non-relevant features. This feature extraction also enhances the model's classification accuracy. Finally, recommendations of a valid product to the users from the extracted features are done by the collaborative Bi-LSTM classifiers. Bi-LSTM is useful, especially for processing sequences of data. Considering both past and future contexts during prediction is of great value in understanding user behavior and preference. This helps in efficient recommendation of the products to the users. The Bi-LSTM classifier is tuned using the Passer learning optimization, created using two common optimization techniques-teaching learning-based optimization (TLBO) [19], Sparrow search algorithm (SSA) [20] [1], improving the performance of the Bi-LSTM classifier. The architecture of the proposed recommendation system in an e-commerce model is shown in Figure 1.

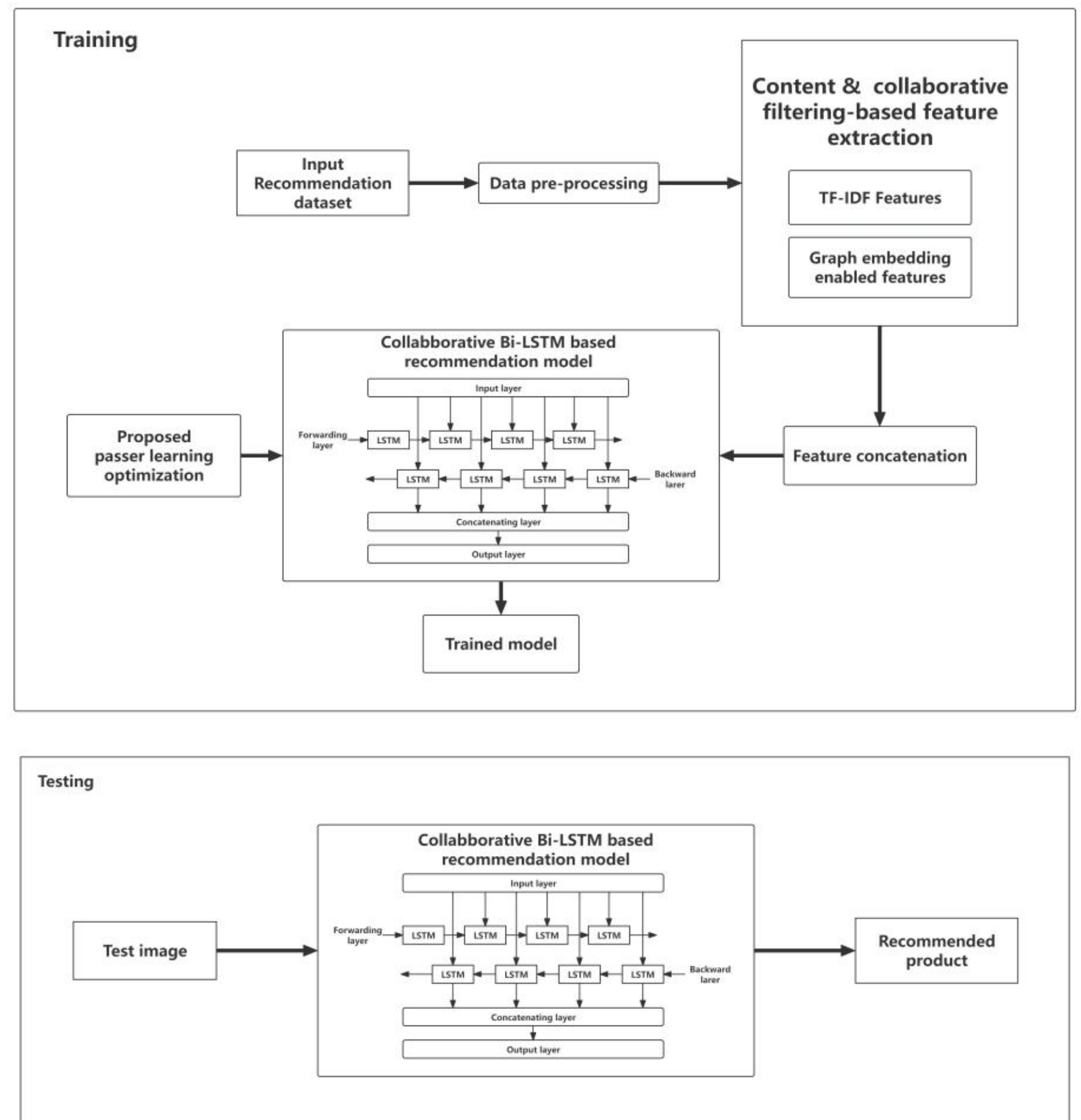

**Fig1.** *Architecture of the proposed recommendation system in the e-commerce model*

### 3.2 Experimental Setup

The research examines how well the proposed PL-optimized Bi-LSTM performs vis-a-vis existing approaches for e-commerce recommendations. For the evaluation, a Python program in the PYCHARM 2022.2.3 tool running on a Windows 10 computer with 8 GB of RAM is used where the GPU is present.

### 3.3 Dataset description (Product recommendation dataset)

The recommendation collection in this research comprises product reviews and associated metadata sourced from Amazon [31]. These reviews are extensive, totaling 143.7 million, and span a period from May 1996 to July 2014. The dataset includes not only reviews but also product metadata and links. The research focuses on three specific datasets: the baby dataset containing 160, 792 reviews; the digital music dataset containing 64, 706 reviews; and the patio lawn garden dataset containing 13, 272 reviews. These datasets form the basis of the analysis and recommendations in this research. The input for the recommendation model is mathematically formulated as the following equation:

$$R = \sum_{i=1}^{a} R_i \tag{1}$$

where, the recommendation dataset is denoted as$R$, and$R_i$ denotes images present in the dataset, ranging from 1 to $a$.

### 3.4 Data Pre-processing

This research uses stemming, lemmatization, and tokenization processes for pre-processing.

***Tokenization*** is the process of splitting text into individual words or tokens, making it easier for computers to understand and analyze the content. This essential step in natural language processing breaks down complex sentences or paragraphs into manageable units for subsequent analysis.

**Stemming** further enhances text processing by removing the last few characters from a word, often leading to incorrect meanings and spelling.

**Lemmatization** also enhances text processing by reducing words to their root or base form, ensuring consistency in word representations and aiding in identifying common semantic meanings within the text data. The pre-processed data is mathematically represented as

$$R = \sum_{i=1}^{a} R_i^* \tag{2}$$

where the pre-processed image is denoted as$R_i^*$.

#### 3.4.1 Content &Collaborative Filtering-based Feature Extraction

Eliminating unnecessary features is an essential phase of feature selection, which avoids overfitting problems by lowering the size of the features. To accurately categorize the polarity, the algorithm is trained for a certain feature that is used to represent the data. The class attribute is thus represented in the smaller feature space by feature selection, which chooses the fewest significant characteristics. The techniques for feature selection can considerably increase classification accuracy and provide users with better knowledge of key class traits, helping to interpret the data. The two steps used in this research to extract TF-IDF features and graph embedding-enabled features are discussed in depth below.

***TF-IDF features:*** The term ‘frequency-inverse document frequency’ (TF-IDF) is one of the most popular weighting metrics for detecting the link between words and documents, commonly used in the process of collecting word features. We can quantify the importance of every phrase in a document by assigning it a numerical value with the aid of TF-IDF. The inverse document rate, or IDF, quantifies the frequency of words appearing across the whole corpus. A term appearing only occasionally in the corpus can still be considered a feature because it has great differentiation potential (Hong & Jung, 2018). As TF-IDF merely computes the importance of each word in a document or collection of documents rather than evaluating each document's entire content, search engines can rank texts with less computational effort. The equation for the TF-mathematical IDF is calculated as follows:

$$Vt_{i,j} = \frac{O_{i,j}}{\sum_a O_{a,j}} \tag{3}$$

$$idt_i = log\frac{|Q|}{|\{j: w_i \in v_j\}|} \tag{4}$$

The denominator of the formula (3) is the total number of times all the words appear in the text, and the value in the numerator of the formula is the number of times each word appears in the document. Formula (4) $|Q|$ shows the overall number of documents in the corpus or the entire number of documents comprising the phrases. If the feature word's $TF - IDFt_1$ value equals the word's value when $Vt_{i,j}$ is multiplied by$idt_i$, the weight of the feature word can also be determined using the formula (5).

$$TF - IDFt_1 = vt_{i,j} \times idt_i \tag{5}$$

***Graph embedding-enabled features*****:** Graph embedding refers to the processes that convert property graphs into a vector or a collection of vectors. Graph topology, vertex-to-vertex relationships, and other pertinent details about graphs, subgraphs, and vertices should all be captured during embedding.

As graph embedding techniques compress every node's attributes in a vector with a smaller dimension, node similarity in the original complicated irregular spaces may be readily measured in the embedded vector spaces using common metrics.

The benefits of graphical representation include time-saving and easier understanding of data. Graph embedding is a dimensionality reduction technique that extracts the most meaningful information from graph data while significantly lowering the computational burden and complexity accompanying large dimensions. This graph embedding feature for datasets 1, 2, and 3 is shown in Figure 2.

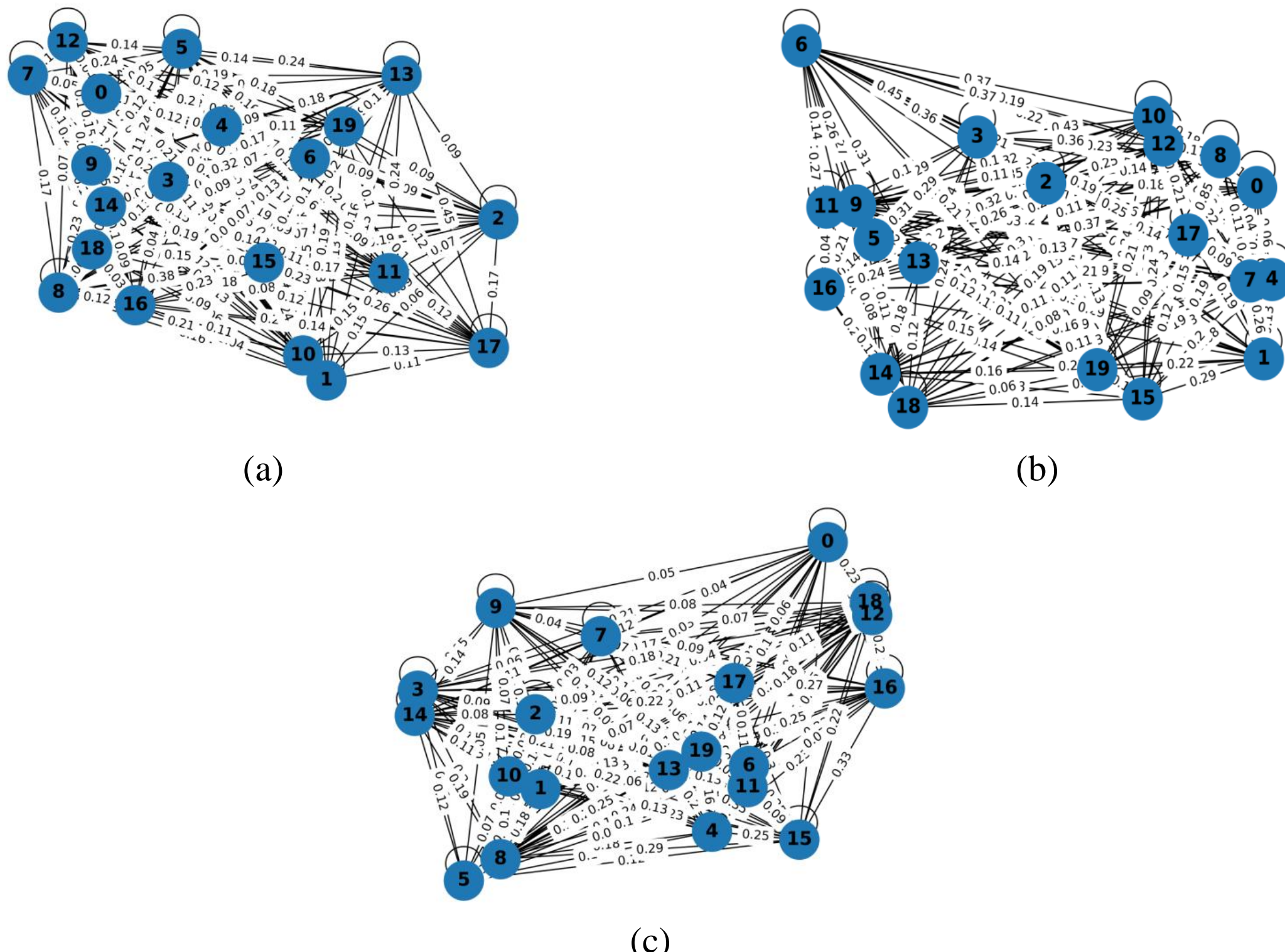

**Fig 2.** *Graph embedding-based features for (a) The baby dataset , (b) The digital music dataset and (c) The patio lawn garden dataset*

**Feature concatenation:** After numerous features with the same dimensions are extracted to the Bi-LSTM classifier for analysis, they are merged using the feature concatenation approach. The various original information is significantly conserved throughout the concatenation method due to the integrated perspective representation of all the concatenated features. The Passer learning optimization-based Bi-LSTM classifier receives the output of feature concatenation and improves the performance during the recommendation system. The optimization method discovers the most effective solution while the overall number of parameters in the features is minimized. Consequently, the model's performance is assessed and trained, using both the test and trained data as modeling input. A training dataset is used to train the model with the dimension of $1 \times 2000$, and the model's accuracy determines the test data.

**3.4.2 Collaborative Bi-LSTM-based Recommendation Model**

Bi-LSTM plays an important role in the recommendation system by making personalized product recommendations for users with the help of sequential information and context. It considers content features TF-IDF, graph embedding, and collaborative information for making informed suggestions. Adding Passer learning optimization further refines the performance of Bi-LSTM, making the recommendations quite accurate and effective. After extraction, the features are fed to collaborative Bi-LSTM classifiers that shall wield the power of Bidirectional Long Short-Term Memory for recommendations. Bi-LSTM is particularly helpful in handling the processing of sequences-be it in NLP tasks or sequential recommendation. It offers an ability to predict, considering past and future contexts-a useful insight for understanding user behaviors and preferences. The performance of the Bi-LSTM classifier is further improved by the system using Passer Learning Optimization. Passer learning optimization is an algorithm that combines the characteristics of a teacher-student relationship inspired by sparrow behavior and is designed to improve convergence rates and performance. It fine-tunes the parameters of the Bi-LSTM classifier to make more accurate recommendations [1].

Figure 3 displays the Bi-LSTM classifier recommendation system. This architecture is comprised of input layers, forwarding layers, backward layers, totally connected layers, and output layers. The Bi-LSTM classifier effectively manages the error gradient by making gates available and providing long-term dependencies, significantly reducing the vanishing gradient problem. The equation demonstrates how the Bi-LSTM classifier is conceptualized mathematically.

$$B_X = A(w_c . x_Y + O_Y . c_{Y-1} + b_c) \tag{6}$$

The features expressed as $R^*_{i,f}$ are carried by the current word embedding represented by $B_X$. The weights for the Bi-LSTM classifier are given by $w_c$ and $O_Y$, whereas $A(X)$ stands for the nonlinear function, $b_c$ determines the bias and $c_{Y-1}$ signifies the hidden state. The classifier's weights and biases are optimized using the Passer learning optimization, which also successfully adjusts the hyperparameters.

$$E_Y = \sigma(w_E . y_Y + O_E . c_{Y-1} + b_E) \tag{7}$$

$$J_Y = \sigma\left(w_J . y_Y + O_J . c_{Y-1} + b_J\right) \tag{8}$$

$$P_Y = \sigma(w_P . y_Y + O_P . c_{Y-1} + c_P) \quad (9)$$

$$m_Y = E_Y . m_{Y-1} + J_Y . tanh(w_m . x_Y + O_m . c_{Y-1} + c_m) \quad (10)$$

$$c_Y = P_Y . tanh(m_X) \quad (11)$$

Here, $E_Y$,$J_Y$ and $P_Y$ stand for the input gate, forget gate, and memory cell, respectively. $m_Y$stands for the memory cell, $P_Y$ for the Hadamard product, and $\sigma$ for the sigmoid function. While the forget gate assists in forgetting previous knowledge, the memory cell in the input gate keeps the currently important information. The output gate determines which information is presented in the internal memory cell and enables many retrievals of crucial information.

**Fig 3.** *Bi-LSTM architecture*

### 3.5 Proposed Passer Learning Optimization

To make the Bi-LSTM classifier work effectively, an effective passer learning optimization should be enabled which can tune these parameters.Thus, the sparrow [19] and teacher [20], which are basically developed with built-in automatic learning features, should be combined with the passer to increase speed, convergence rate, and improvement in performance order. A basic explanation of an algorithm along with a mathematical notational description is furnished further in the subsequent sections.

***Motivation:*** PLO is the hybrid algorithm proposed in this work, which incorporates the teaching capabilities of a "teacher" with inspiration from the sparrow behavior of the Sparrow search algorithm to enhance the performance and convergence rate. The sparrow is believed to be one of the most intelligent birds, classified into "producers", "scroungers", and "investigators." Producers have energy reserves that help lead scavengers to the food, while scroungers follow the best food providers. The algorithm also implements a security mechanism whereby sparrows chirp to raise an alarm, and producers lead the scavengers to safety if the alarm threshold is violated. This strategy has been adopted from the foraging strategy of sparrows. TLBO is a population-based algorithm that draws inspiration from classroom dynamics. It has two phases: the teacher phase and the learner phase. In the teacher phase, students learn from a knowledgeable teacher and try to improve their knowledge. In the learner phase, students share and improve their knowledge through interactions such as group discussions and learning from more knowledgeable peers. TLBO emulates the teaching-learning process in a classroom environment. These algorithms, PL and TLBO, draw from the natural behaviors of sparrows and classroom dynamics in order to enhance convergence and learning within optimization problems.

***Mathematical model of the proposed Passer learning optimization*:** The producer, scrounger, and investigator are the three categories used to represent the mathematical model of the proposed Passer learning optimization, and those behaviors are mathematically formulated in the sections below.

**3.5.1 Initialization**

The solutions of the developed optimization are randomly initialized with the lower bound $(lb)$ and upper bound$(ub)$. The initialized solutions are presented as,

$$x_{ij}^{t} = rand \cdot (ub - lb) + lb \tag{12}$$

where, $rand$ is the random number and $x_{ij}^{t}$ is the solution position at iteration $t$.

**3.5.2 Evaluation of fitness**

The optimal solution establishes the fitness of the developed optimization. The solution with the minimum mean square error (MSE) is the best fitness and can be stated as,

$$fit(x_{ij}^{t}) = \min\left(MSE(x_{ij}^{t})\right) \tag{13}$$

**3.5.3 Solution update**

After determining the fitness functions, the solutions are updated using different phases are presented below,

***3.5.3.1Position Update of Producer***

Producers in Sparrow are responsible for finding food and controlling the movement of the entire population, though producers with higher fitness ratings have priority when it comes to finding food. Consequently, the producers have access to a greater range of food sources than the scroungers, and the teachers' teaching characteristics improve the producers' capability to explore more search areas. When the sparrows see a predator, they immediately begin chirping as an alert. If the alarm value beats the safety level, the producers are required to direct all scavengers to a safe area. The location update producer's mathematical formula is as follows:

$$x_{ij}^{t+1} = \begin{cases} x_{ij}^{t} \cdot ex\left(\dfrac{-h}{\alpha_1 \cdot t_{max}}\right) + \left(M_{ij}^{t} - x_{ij}^{t}\right)R_2 < AP \\ \left(x_{ij}^{t} + r_i * fl\right) \times \left(M_{ij}^{t} - x_{ij}^{t}\right)R_2 \geq AP \end{cases} \tag{15}$$

The current iteration is denoted as $h$, the random number as $\alpha_1$ which ranges from 0 to 1, the flight length as $fl$, the uniform random number as $r_i$ which ranges from 0 to 1, the maximum number of iterations as $M$, the alarm value as $R_2$, and awareness probability as $AP$. $x_{ij}^{t}$ represents the current position of the$j^{th}$dimension of the $i^{th}$sparrow at the iteration $t$. When $R_2 < AP$it indicates that there are no predators nearby. All sparrows must immediately fly to other safe regions if $R_2 \geq AP$, this indicates that some sparrows have identified the predator.

***3.5.3.2Position Update of Investigators***

In the sparrow system, investigators are randomly selected from the population. They emit signals to prompt sparrows to move safely when predators enter the area. However, the sparrow's automatic prediction capacity is not properly interpreted, leading to a slower performance and poor convergence rate. To improve convergence rate and performance, the automatic learning capability of the teacher learning system is integrated with the sparrow's abilities, enabling independent predator identification. This integration enhances global convergence and minimizes time complexity. The mathematical formula for the combined algorithm is as follows:

$$x_{ij}^{t+1} = |x| \cdot A^{+} \cdot L - \alpha M_{ij}^{t} \tag{16}$$

$$x_{ij}^{t+1} = \alpha x_{ij}^{t} - \frac{1}{2}\alpha x_{ij}^{t-1} + |A^{+} \cdot L - \alpha M_{ij}^{t}| \tag{17}$$

where $A$ stands for a $1 \times d$matrix, each of whose elements is given a random value of 1 or 1, and $A^{+} = AT(AAT)^{-1}$. $L$ displays a matrix of $1 \times d$whose elements are all 1.

### 3.5.4 Position Update of Scrounger

Some scavengers keep a close eye on the producers and engage in feeding competition with them to increase the pace of their own predation, and the teaching traits of the teachers make the scavengers move to the next location with its capabilities of improving the means of the class. As soon as they discover that the producer has located delicious food, they promptly leave their current location to compete for food. If they are successful, they can get the food right away; if not, the process continues. The scrounger's position update formula is as follows:

$$x_{ij}^{t+1} = \begin{cases} Q \cdot ex\left(\dfrac{M_{ij}^{t-1} - x_{ij}^{t}}{i^2}\right) + rand\left(\dfrac{x_p^t - x_{ij}^t}{f\left(x_{ij}^{t+1}\right)}\right) \\ x_{ij}^t + k\left(\dfrac{x_{ij}^t - M_{ij}^t}{f\left(x_{ij}^t\right) - f\left(x_{ij}^{t-1}\right)}\right) + Q \cdot v_{ij}^t \end{cases} \tag{18}$$

where $Q$ is a random number that follows the normal distribution, the producer's best solution is denoted as $x_p^t$, the $i^{th}$ sparrow's velocity in the $j^{th}$ dimension at iteration $t$ is denoted as $v_{ij}^t$, the sparrow's direction of motion and the step size control coefficient are both denoted as $k$and the fitness functions of the current, next, and previous iterations are denoted as $f\left(x_{ij}^t\right)f\left(x_{ij}^{t+1}\right)$and $f\left(x_{ij}^{t-1}\right)$.

### 4.5.5 Declaration of best solution

The best solution must be determined by re-evaluating the solutions' fitness values once the solutions have been updated.

### 3.5.6 Termination

The optimization process iterates further, validating the condition $(t < t_{max})$ and re-evaluating fitness to get the optimum solution. The Pseudo code for the proposed Passer learning optimization is provided in Algorithm 1.

**Algorithm 1:** Pseudo code for the proposed Passer learning optimization

| **S.No** | **Pseudo code for the proposed Passer learning optimization** |
|---|---|
| 1. | Initialization |
| 2. | $M$: maximum number of iterations |
| 3. | $R_2$: alert value |
| 4. | $Q$: random value |
| 5. | Initialize population |
| 6. | $t = 1$ |
| 7. | While $(t < t_{max})$ |
| 8. | Update the position of the producer |

9. $$x_{ij}^{t+1} = \begin{cases} x_{ij}^{t} \cdot ex\left(\frac{-h}{\alpha_1 \cdot M}\right) + \left(M_{ij}^{t} - x_{ij}^{t}\right)R_2 < AP \\ \left(x_{ij}^{t} + r_i * fl\right) \times \left(M_{ij}^{t} - x_{ij}^{t}\right)R_2 \geq AP \end{cases}$$

10. $$R_2 = rand(1)$$

11. $$for(i = 1)$$

12. Update the position of the scrounger

13. $$x_{ij}^{t+1} = \begin{cases} Q \cdot ex\left(\frac{M_{ij}^{t-1} - x_{ij}^{t}}{i^2}\right) + rand\left(\frac{x_p^t - x_{ij}^t}{f\left(x_{ij}^{t+1}\right)}\right) \\ x_{ij}^{t} + k\left(\frac{x_{ij}^{t} - M_{ij}^{t}}{f\left(x_{ij}^{t}\right) - f\left(x_{ij}^{t-1}\right)}\right) + Q \cdot v_{ij}^{t} \end{cases}$$

14. End for

15. $$for(i = N + 1)$$

16. Update the position of the investigators

17. $$x_{ij}^{t+1} = \alpha x_{ij}^{t} - \frac{1}{2}\alpha x_{ij}^{t-1} + \left|A^{+} \cdot L - \alpha M_{ij}^{t}\right|$$

18. End for

19. Re-evaluate the fitness

20. Declare the best solution

21. End while

---

### 3.6 Performance Metrics

The performance of the proposed method is measured using the following metrics.

***Precision***: Precision $C_{precision}$ is defined as the ratio of correctly predicted positive results to all positively predicted by the PL-optimized Bi-LSTM.

$$C_{pre} = \frac{T_r P}{T_r P + F_a P} \tag{19}$$

where $T_r P$represents the true positive, $F_a P$ the false positive and $F_a N$ the false negative.

*Recall:*Recall $C_{recall}$ is calculated by dividing the sum of true positives and false negatives by the total number of true positives.

$$C_{recall} = \frac{T_r P}{T_r P + F_a N} \tag{20}$$

*F1 score* is the harmonic mean between recall and precision and is represented as follows:

$$C_{f1-score} = 2\frac{C_{pre} \times C_{recall}}{C_{pre} + C_{recall}} \tag{21}$$

*Mean Squared Error (MSE):*When comparing real and estimated data, the difference is known as mean square error.

## 4. Results

The results derived by employing the PL-optimized Bi-LSTM for the e-commerce recommendation system are discussed below.

### 4.1 Performance Evaluation

For testing the classifier's performance during different epochs, the PL-optimized Bi-LSTM classifier is put through a performance evaluation, as detailed below.

The consolidated performance evaluation table presents the efficacy of PL-optimized Bi-LSTM on three datasets: baby, digital music, and patio lawn garden, across different training percentages like 40%, 50%, 60%, 70%, 80%, and 90%. In the baby dataset, at a 90% training percentage, the maximum F1 score and MSE were recorded as 88.58% and 1.24%, respectively, indicating very good predictive accuracy with low error rates. Also, the digital music dataset is performing strongly at 90% training with an F1 score of 88.46% and lowest MSE of 0.48%. Patio lawn garden dataset proves to be tough, hence the model holds robust in diversity, reaching up to an F1 score of 92.51% with an MSE of 1.58% at maximum training. The precision and recall metrics across these datasets are seen to increase consistently with increasing training percentages, hence showing the capability of the model to learn effectively and adapt to various types of e-commerce data, hence ensuring highly accurate recommendations. This table shows the effectiveness and adaptability of the PL-optimized Bi-LSTM in handling complex recommendation tasks across different domains, as shown in Table 1.

Table 1: Consolidated Performance Evaluation of the PL-Optimized Bi-LSTM on Multiple Datasets

| Dataset | Metric | 40% | 50% | 60% | 70% | 80% | 90% |
|---|---|---|---|---|---|---|---|
| **Baby Dataset** | F1 Score | 66.43% | 73.83% | 75.94% | 78.18% | 88.58% | 88.58% |
| | MSE | 1.66% | 1.62% | 1.54% | 1.36% | 1.24% | 1.24% |
| | Precision | 66.74% | 75.07% | 77.89% | 79.43% | 92.69% | 92.69% |
| | Recall | 56.31% | 79.74% | 80.93% | 90.93% | 92.69% | 92.69% |
| **Digital Music Dataset** | F1 Score | 80.72% | 83.21% | 83.66% | 86.55% | 88.46% | 88.46% |
| | MSE | 0.61% | 0.60% | 0.53% | 0.50% | 0.48% | 0.48% |
| | Precision | 74.86% | 80.61% | 85.12% | 86.97% | 92.43% | 92.43% |
| | Recall | 75.83% | 90.26% | 90.47% | 91.74% | 93.47% | 93.47% |
| **Patio Lawn Garden Dataset** | F1 Score | 89.15% | 89.23% | 90.35% | 91.91% | 92.51% | 92.51% |
| | MSE | 2.32% | 2.16% | 2.08% | 1.77% | 1.58% | 1.58% |
| | Precision | 82.77% | 85.69% | 86.67% | 89.95% | 91.90% | 91.90% |
| | Recall | 60.54% | 74.95% | 75.52% | 78.96% | 90.76% | 90.76% |

### 4.2 Comparative Methods

Table 2 shows the comparative performance, representing the efficiency of the PL-optimized Bi-LSTM classifier against other benchmark models comprising HPMFM, SCCF, A3NCF, simple BiLSTM, BiLSTM with Harris Hawk Optimization, and BiLSTM with Particle Swarm Optimization on three datasets: baby, digital music, and patio lawn garden. PL-optimized Bi-LSTM outperforms all the mentioned models in F1 Score, MSE,

Precision, and Recall at a 90% training percentage for all the above-mentioned metrics. Most noticeably, it reached the highest F1 Scores, 90.72%, 90.43%, and 94.24%, respectively, which shows the excellence in the precision of recommendation predictions. The model also shows the lowest MSE rates, thus showing its preciseness in modeling with minimized errors. Precision and Recall metrics confirm its robustness, with scores considerably higher than those from competing methods, thus illustrating its ability to provide highly relevant and accurate recommendations across diverse e-commerce settings. This comparative analysis has indicated the advanced capability of PL-optimized Bi-LSTM in handling complex recommendation tasks with better results than the conventional and other optimized LSTM models.

Table 2: Comparative Performance of PL-Optimized Bi-LSTM and Other Models

| **Dataset** | **Metric** | **HPMFM** | **SCCF** | **A3NCF** | **BiLSTM** | **BiLSTM with HHO** | **BiLSTM with PSO** | **PL-Optimized Bi-LSTM** |
|---|---|---|---|---|---|---|---|---|
| **Baby Dataset** | F1 Score | 66.91% | 68.25% | 68.44% | 86.25% | 86.43% | 86.87% | **90.72%** |
| | MSE | 2.52% | 2.47% | 2.42% | 1.34% | 1.25% | 1.24% | **1.22%** |
| | Precision | 57.66% | 57.89% | 58.81% | 79.90% | 82.25% | 83.19% | **93.52%** |
| | Recall | 74.70% | 75.64% | 83.37% | 83.83% | 83.89% | 84.10% | **92.54%** |
| **Digital Music Dataset** | F1 Score | 56.71% | 59.07% | 60.25% | 80.95% | 83.86% | 88.93% | **90.43%** |
| | MSE | 1.80% | 1.74% | 1.71% | 0.77% | 0.63% | 0.59% | **0.53%** |
| | Precision | 52.63% | 56.61% | 57.74% | 85.83% | 87.55% | 92.10% | **95.00%** |
| | Recall | 61.44% | 62.67% | 66.16% | 75.57% | 86.33% | 88.78% | **94.20%** |
| **Patio Lawn Garden Dataset** | F1 Score | 61.71% | 62.21% | 67.27% | 79.55% | 83.83% | 88.67% | **94.24%** |
| | MSE | 2.41% | 2.36% | 2.05% | 1.73% | 1.62% | 1.47% | **1.41%** |
| | Precision | 53.59% | 54.66% | 60.00% | 74.69% | 78.81% | 90.69% | **96.12%** |
| | Recall | 56.18% | 57.31% | 59.06% | 73.10% | 80.60% | 92.00% | **93.38%** |

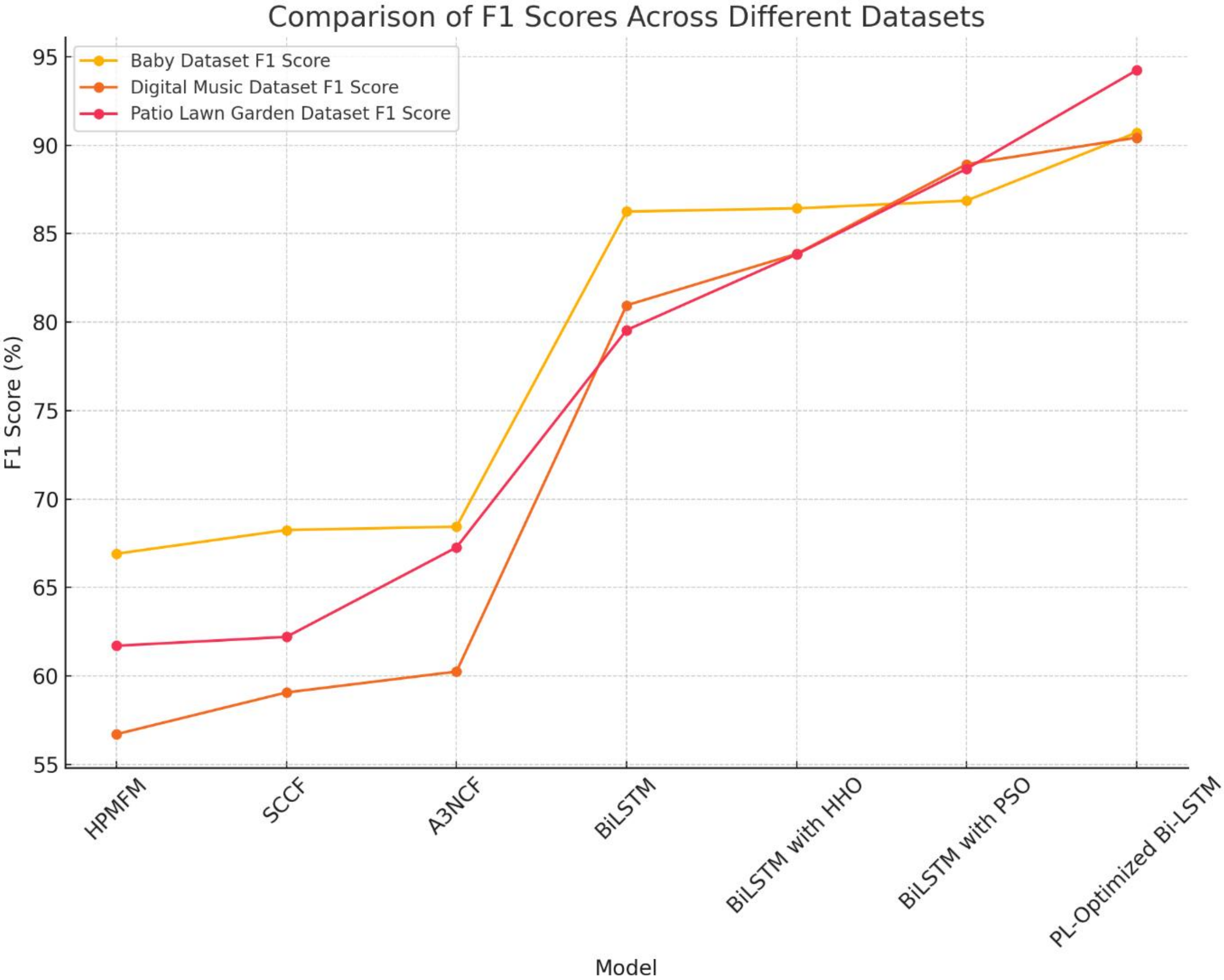

**Fig 4.** Comparative Performance of PL-Optimized Bi-LSTM and Other Models

## 5. Comparative discussion

The comparison indeed highlights the strengths of the PL-optimized Bi-LSTM classifier against the traditional and other LSTM-enhanced models along many dimensions of performance. Our results support the aim of the research in improving the accuracy and robustness of e-commerce recommendation systems.

Accuracy and Efficiency: PL-optimized Bi-LSTM has obtained the highest F1 Scores and the lowest MSE across all datasets evaluated, namely, baby, digital music, and patio lawn garden, hence proving to be the most predictive with efficiency. These improvements can be seen in the visualizations of the baby dataset, where the PL-optimized Bi-LSTM achieved a 90.72% F1 Score, with an MSE of 1.22%, against competing models like BiLSTM, which used Particle Swarm Optimization to give 86.87% for F1 Score at 1.24% MSE. These metrics are of great importance because they are a direct reflection of the ability of the system to minimize errors while it works towards higher predictive accuracy; hence, an improved performance toward the goal of optimization for e-commerce recommendation systems.

Robustness in Cold-Start Scenarios: Of importance is the robustness of PL-optimized Bi-LSTM when considering cold-start scenarios that usually see most of the conventional systems fall. This is reflected in the huge improvements of recall metrics across the board. For example, in the patio lawn and garden dataset, the recall rose from 92.00% in the BiLSTM with PSO to 93.38% in the PL-optimized Bi-LSTM. The improved recall reflects that the model was able to identify relevant products with minimal user interaction data, a common problem with recommendation systems.

`Relevance to Research Objectives: The results contribute directly to the major research objectives, as it will be shown in this section that the PL-optimized Bi-LSTM enhances not only the accuracy and efficiency of the recommendations but also scalability and challenging cold-start situations. By incorporating state-of-the-art graph embedding techniques and Passer Learning Optimization, the model responds well to the need for developing a scalable, accurate, and robust system that can operate effectively in diverse dynamic e-commerce environments..

**Table 3.** *Comparative discussion of the proposed Pl-optimized bi-LSTMmethod using training percentage-90*

| Dataset | Methods/Metrics | HPMFM | SCCF | A3NCF | BiLSTM | BiLSTM with HHO | BiLSTM with PSO | Proposed Method |
|---|---|---|---|---|---|---|---|---|
| **The baby dataset** | **F1-score** | 66.91 | 68.25 | 68.44 | 86.25 | 86.43 | 86.87 | 90.72 |
| | **MSE** | 2.52 | 2.47 | 2.42 | 1.34 | 1.25 | 1.24 | 1.22 |
| | **Precision** | 57.66 | 57.89 | 58.81 | 79.90 | 82.25 | 83.19 | 93.52 |
| | **Recall** | 74.70 | 75.64 | 83.37 | 83.83 | 83.89 | 84.10 | 92.54 |
| **The digital music dataset** | **F1-score** | 56.71 | 59.07 | 60.25 | 80.95 | 83.86 | 88.93 | 90.43 |
| | **MSE** | 1.80 | 1.74 | 1.71 | 0.77 | 0.63 | 0.59 | 0.53 |
| | **Precision** | 52.63 | 56.61 | 57.74 | 85.83 | 87.55 | 92.10 | 95.00 |
| | **Recall** | 61.44 | 62.67 | 66.16 | 75.57 | 86.33 | 88.78 | 94.20 |
| **The patio** | **F1-score** | 61.71 | 62.21 | 67.27 | 79.55 | 83.83 | 88.67 | 94.24 |
| | **MSE** | 2.41 | 2.36 | 2.05 | 1.73 | 1.62 | 1.47 | 1.41 |
| **lawn garden dataset** | **Precision** | 53.59 | 54.66 | 60.00 | 74.69 | 78.81 | 90.69 | 96.12 |
| | **Recall** | 56.18 | 57.31 | 59.06 | 73.10 | 80.60 | 92.00 | 93.38 |

### 5.1 Computational Analysis

Weight and bias are the hyperparameters present in the Bi-LSTM classifier that are finely tuned using the proposed PLO algorithm, which helps processing attain the optimal solution with a minimum run leading to a high convergence rate; this is visually interpreted in Figure 5 and Table 4. The study includes specific settings and parameters used in the final run of the algorithm, such as 50 epochs, a learning rate of 0.001, a batch size of 64 and input dimensions of (None, 1) and (None, 100) and an output dimension of (None, 1).

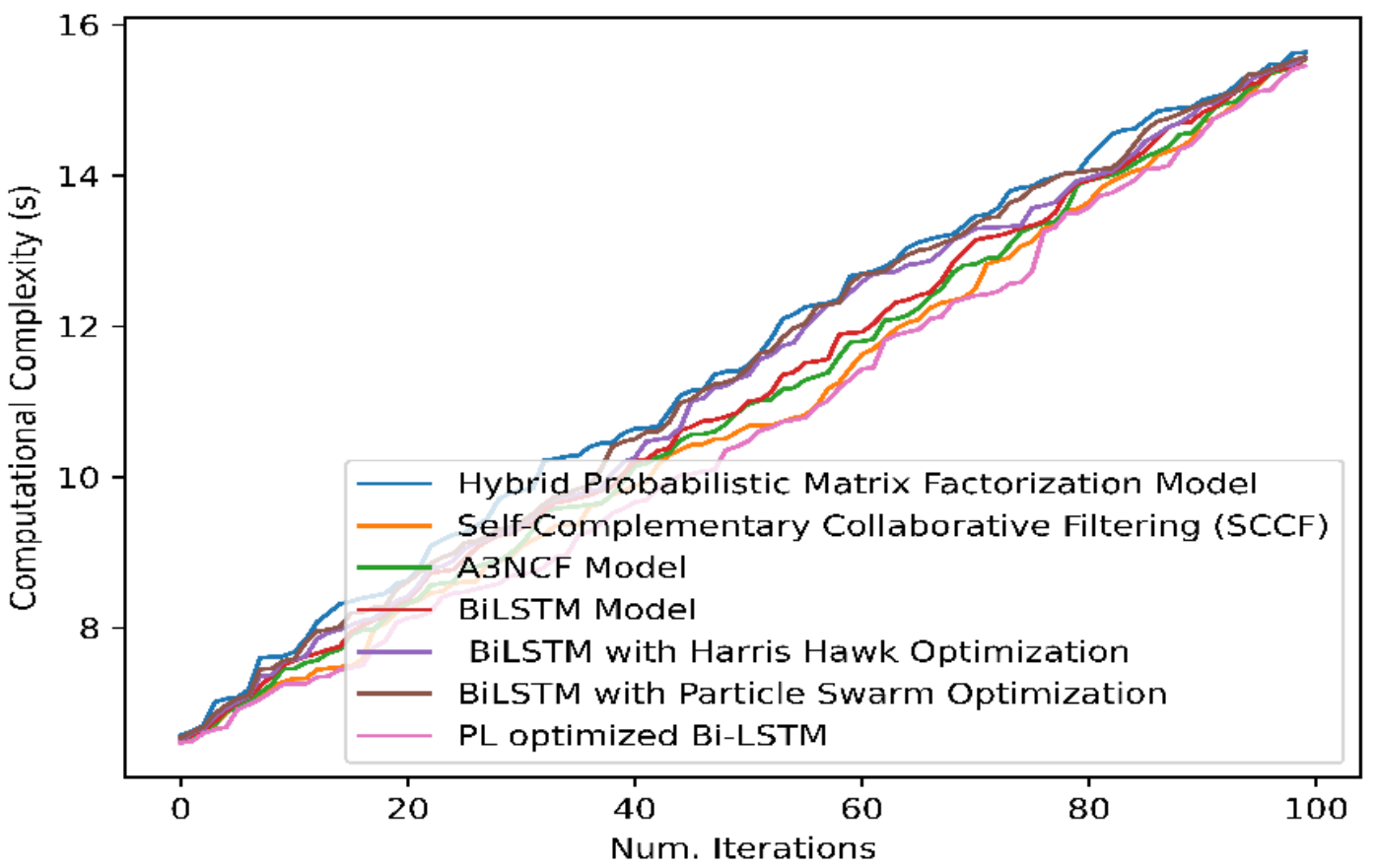

**Fig 5.***Computational analysis of PL-optimized Bi-LSTM Model*

**Table 4.** *Computational Analysis*

| Methods | Iteration-60Time (ms) | Iteration-90Time (ms) |
|---|---|---|
| HPMFM | 12.696 | 15.003 |
| SCCF | 11.627 | 14.716 |
| A3NCF | 11.800 | 14.718 |
| BiLSTM | 11.923 | 14.837 |
| BiLSTM with HHO | 12.589 | 14.938 |
| BiLSTM with PSO | 12.686 | 14.951 |
| PL-optimized Bi-LSTM | 11.434 | 14.574 |

### 6. Conclusion and Future Work

In an e-commerce personalized recommendation system model, time, cost constraints, and slower convergence affect the working. Hence, this research has established that the PL-optimized Bi-LSTM classifier significantly enhances e-commerce recommendation systems, achieving up to 94.24% F1 score and reducing MSE as low as 1.22%. The results not only show improved accuracy and efficiency but also reveal the model's capability to handle diverse e-commerce environments.

Future Directions: This can be extended further by incorporating the PL-optimized Bi-LSTM with GNNs in future research for enhanced feature extraction capabilities. The model can further be optimized regarding adaptiveness in learning rates and can be used for cross-domain recommendation systems for more efficient real-time responsiveness and accuracy. These kinds of hybrid approaches are going to tune up recommendation systems for much better user experience and computational efficiency.

**Acknowledgements**

The authors gratefully acknowledge the financial support from Wenzhou-Kean University.

**Availability of data and materials**

All data based on references

**Funding**

This work was supported by Leading Talents of Provincial Colleges and Universities, Zhejiang-China (Grant No: KY20220214000024) and General Program - Education Department of Zhejiang Province (Grant No. Y202045131).

**Ethics declarations:**

**Ethics approval and consent to participate**

Not applicable.

**Consent for publication**

The authors give the Publisher permission to publish the work.

**Competing interests**

The authors declare that they have no competing interests.